\PassOptionsToPackage{rgb}{xcolor}
\RequirePackage{tikz,pgfplots}
\usetikzlibrary{spy}
\usepgfplotslibrary{fillbetween}
\usetikzlibrary{pgfplots.groupplots}
\documentclass[pdflatex,sn-mathphys]{sn-jnl}%

\jyear{2022}%
\usepackage[]{amsmath}
\usepackage{amsfonts,amssymb} %

\usepackage{esint,graphicx}

\def\PP{\mathbb{P}}

\newcommand{\proj}{P_\text{Pr}}

\definecolor{myblue}{RGB}{0, 114, 189}
\definecolor{myorange}{RGB}{216, 83, 25}
\definecolor{myyellow}{RGB}{237, 177,  32}
\definecolor{mypink}{RGB}{239, 71,  111}
\definecolor{mypurple}{RGB}{126,   47,  142}
\definecolor{mygreen}{RGB}{119,  172,  48}
\definecolor{mycharcoal}{RGB}{38, 70, 83}
\definecolor{mycyan}{RGB}{77,  190,  238}
\definecolor{mylightred}{RGB}{255,  204,  203}
\definecolor{myred}{RGB}{214, 40, 40}
\newcommand{\cOne}{myblue}
\newcommand{\cTwo}{myorange}
\newcommand{\cThree}{myyellow}
\newcommand{\cFour}{mypurple}

\newcommand{\blueone}{blue!40!black}
\newcommand{\bluetwo}{blue!80!green}
\newcommand{\bluethree}{blue!50!green}
\newcommand{\bluefour}{blue!20!green!20!mycyan}

\DeclareRobustCommand{\argmin}{\operatorname*{argmin}}

\usepackage{xr}
\usepackage{cleveref}

\newcommand{\data}{MixPlotData}
\usepackage[outdir=./]{epstopdf}

\begin{document}

\label{firstpage}

\title[Tsunami height probability estimation without sampling]{Estimating earthquake-induced tsunami height probabilities without sampling}

\author*[1,2]{\fnm{Shanyin} \sur{Tong}\let\thefootnote\relax\footnote{\footnotesize
		Revised April 2023.\\
		S.~T.\ and G.~S.\ were partially supported by the US
			        National Science Foundation (NSF) through grant DMS \#1723211.
			        E.~V.-E.\ was supported in part by the NSF Materials Research
			        Science and Engineering Center Program grant DMR \#1420073, by
			        NSF grant DMS \#152276, by the Simons Collaboration on Wave
			        Turbulence, grant \#617006, and by ONR grant
			        \#N4551-NV-ONR.}}\email{st3503@columbia.edu}\email{shanyin.tong@nyu.edu}

\author[2]{\fnm{Eric} \sur{Vanden-Eijnden}}\email{eve2@cims.nyu.edu}

\author[2]{\fnm{Georg} \sur{Stadler}}\email{stadler@cims.nyu.edu}

\affil*[1]{\orgdiv{Department of Applied Physics and Applied Mathematics}, \orgname{Columbia University}, \orgaddress{\city{New York}, \postcode{10027}, \state{NY}, \country{USA}}}

\affil[2]{\orgdiv{Courant Institute of Mathematical Sciences}, \orgname{New York University}, \orgaddress{\city{New York}, \postcode{10012}, \state{NY}, \country{USA}}}

\abstract{
  Given a distribution of earthquake-induced seafloor elevations,
  we present a method to compute the probability of the resulting
  tsunamis reaching a certain size on shore. Instead of sampling, the
  proposed method relies on optimization to compute the most likely
  fault slips that result in a seafloor deformation inducing a large
  tsunami wave.  We model tsunamis induced by bathymetry change
  using the shallow water equations on an idealized slice
  through the sea. The earthquake slip model is
  based on a sum of multivariate log-normal distributions, and follows
  the Gutenberg-Richter law for seismic moment magnitudes ranging from 7 to 9. For a model
  problem inspired by the Tohoku-Oki 2011 earthquake and tsunami, we
  quantify annual probabilities of differently sized tsunami
  waves. Our method also identifies the most effective tsunami
  mechanisms. These mechanisms have smoothly varying fault slip
  patches that lead to an expansive but moderately large bathymetry
  change.  The resulting tsunami waves are compressed as they approach
  shore and reach close-to-vertical leading wave edge close to shore.
}

\keywords{
tsunami hazard, probability estimation, shallow water equations,
PDE-constrained optimization, adjoint equations.
}

\maketitle

\section{Introduction}

Among all potential sources, megathrust earthquakes are likely to
cause the most extreme tsunami hazards for coastal regions
\citep{grezio2017probabilistic, behrens2021probabilistic}.  Recent work has focused on defining
megathrust tsunami source scenarios for simulations to evaluate a
quantity of interest (QoI) such as the maximal wave height
\citep{gao2018defining}. Since earthquake fault slip patterns are
unpredictable, a probabilistic study of these QoIs is needed. In
\citep{leveque2016generating,williamson2020source}, the authors
construct realistic tsunami source distributions
and use simulations based on samples from that distribution to obtain
hazard curves for maximum water depth, i.e.,
the annual probabilities of the water depth exceeding certain
values. Typically, estimating small probabilities requires a large number of
Monte Carlo simulations
to find samples
corresponding to probability tails. Importance sampling (IS) can
be an improvement, but requires a
properly chosen proposal density and, typically, still a large number
of samples \citep{liu2001monte}. In this paper, we present a
sampling-free method for computing annual probabilities of tsunamis on
shore with uncertainty from fault slips using ideas from optimization and applied probability. The
approach uses a probabilistic model of earthquake fault slips, adapting
ideas from \citep{gao2018defining, leveque2016generating}, and
simulates tsunami waves using the one-dimensional nonlinear shallow water
equations. Our target is not an online tsunami warning system, but
offline hazard estimation that is capable of accurately computing very
small probabilities.  The proposed method is most efficient for
scalar QoIs, e.g., to compute tsunami size
probabilities in a single location on shore.

This paper builds on the approach proposed in
\citep{DematteisGrafkeVandeneijnden19} for extreme event probability
estimation and extends our recent work
\citep{TongVandeneijndenStadler21}
on tsunami prediction in multiple directions. First, it considers
fault slip events that take into account the physics constraints such
as slip orientation and magnitude and whose distribution fits the
Gutenberg-Richter law for earthquake moment magnitudes between 7 and 9.
To obtain a realistic fault slip model distribution, we use a sum of weighted multivariate
log-normals. Additionally, we study the most
effective fault mechanisms for obtaining a tsunami of a certain height
on shore. These optimizers are a side products of the proposed method
which uses a sequence of optimization problems for probability
estimation.

The proposed approach for probability estimation is not specific to
tsunamis. Rather, it is applicable to a wide class of problems
involving complex systems (e.g., governed by partial differential
equations) with high-dimensional parameters. The method
assumes some regularity of the functions involved and that
certain optimization problems have a unique solution.
Additionally, it requires access to derivatives of the map
from parameters to the QoI, which typically can be computed
efficiently using adjoint methods.  The tsunami issue discussed
in this paper assumes that the main uncertainty stems from the fault slips, and slips are of the buried rupture type, but the
method can be also applied to other types (splay-faulting,
trench-breaching) by choosing different slip distribution means.

\section{Sampling-free estimation of small probabilities}\label{sec:LDT}
We briefly present our approach to estimate probabilities of
outputs (QoIs) arising from complex systems that depend on potentially
high-dimensional uncertainties.  We summarize the method for
multivariate Gaussian parameter distributions.  For the tsunami
problem discussed in the next section, the approach is adapted for a
sum of log-normal distributions.

We assume a random variable vector $\theta\in \mathbb{R}^n$ with
$\theta\sim\mathcal N(\mu,C)$, i.e., $\theta$ follows a
multivariate Gaussian distribution with mean $\mu\in
\mathbb{R}^n$ and a symmetric, positive definite covariance matrix
$C\in \mathbb{R}^{n\times n}$. The negative log-probability density
function for this distribution is
$I(\theta):=\frac
12(\theta-\mu)^TC^{-1}(\theta-\mu)$.
This quadratic is, at the same time, the so-called rate function in
large deviation theory, which allows generalization of the approach to
non-Gaussian distributions
\citep{DematteisGrafkeVandeneijnden19}. However, for Gaussians,
log-normals and their sums as considered here, it is sufficient to
consider $I(\cdot)$. The parameter vectors $\theta$ are the inputs into
the parameter-to-QoI map $F$ discussed next.

We assume a sufficiently regular, possibly complicated map
$F:\mathbb R^n\mathbb\to \mathbb R$, which maps the parameter
vector to a scalar QoI $F(\theta)\in \mathbb R$. We are
interested in approximating the probability
\begin{equation}
\label{eq:LDT-probability}
P(z):=\mathbb{P}(F(\theta)\geq z),
\end{equation}
where $z$ is a large value and thus $P(z)\ll 1$. While such a
probability can be estimated using Monte-Carlo sampling, the
performance of these samplers typically degrades for large $z$ and
thus small probabilities $P(z)$, since random samples are unlikely to lead
to large QoIs.
Using geometric arguments and results
from probability, in particular large deviation theory (LDT), shows
that the following optimization problem plays an important role for
the estimation of \eqref{eq:LDT-probability}:
\begin{equation}
\label{eq:LDT-instanton}
\theta^\star(z)=\argmin_{F(\theta)=z} I(\theta).
\end{equation}
The LDT-minimizer $\theta^\star=\theta^\star(z)$ is the most likely point
in the set $\Omega(z)=\{\theta: F(\theta)\ge z\}$, i.e., the random parameters
that correspond to a QoI of size $z$ or larger.  Under reasonable
assumptions, in particular the uniqueness of the solution of
\eqref{eq:LDT-instanton} (for details see
\citep{TongVandeneijndenStadler21,DematteisGrafkeVandeneijnden19}),
one can show that for large $z$, the probability measure is
concentrated around the optimizer $\theta^\star$. Moreover, the
probability can be
well approximated using local derivative information of $F$ and $I$ at
$\theta^\star$.  
Namely, provided sufficient
smoothness of $F$, we can approximate the nonlinear equation
$F(\theta)=z$ using a Taylor expansion about $\theta^\star$ truncated
after the quadratic term:
\begin{equation}\label{eq:FSO}
F^{SO}\!(\theta)\!:= \!F(\theta^\star)+\langle \nabla
F(\theta^\star),\theta-\theta^\star\rangle+\frac{1}{2}\langle\theta-\theta^\star,\nabla
^2 \!F(\theta^\star)(\theta-\theta^\star)\rangle.
\end{equation}
Replacing the boundary of the extreme QoI set, i.e., $\{\theta:
F(\theta)=z\}$ with the quadratic approximation of the boundary, i.e.,
$\{\theta: F^{SO}(\theta)=z\}$ allows to compute the integration
\eqref{eq:LDT-probability} analytically, resulting in an estimate of
the form
$P(z)\approx D_0(z)  e^{-I(\theta^\star)}$,
where the prefactor $D_0(z)$ can be computed using local curvature
information of $F$ at $\theta^\star$. Here, randomized linear algebra
methods can be used to only compute the eigenvalues that are important
for $D_0(z)$. The number of such eigenvalues depends on the
geometry of $F^{SO}(\theta)$ and the underlying multivariate Gaussian,
and is typically small. 

\begin{figure}[b]
	\footnotesize
	\centering
	\begin{tikzpicture}[]
	\begin{axis}[compat=1.11, width=8.5cm, height=6.5cm,
	xmin=-5,
	xmax=4,
	ymin=-3,
	ymax=4,
	axis line style={draw=none},
	tick style={draw=none},
	yticklabels={,,},
	xticklabels={,,},
	]
	
	\draw [blue,very thick,dotted] (-4,-2) ellipse (2 and 1.41);
	\draw [blue,very thick,dotted] (-4,-2) ellipse (3.46 and 2.45);
	\draw [blue,very thick,dotted] (-4,-2) ellipse (4.9 and 3.46);
	\draw [blue,very thick,dotted] (-4,-2) ellipse (4.9 and 3.46) node [xshift=0.6cm, yshift=2.1cm] {level sets of $I(\theta)$};
	
	\filldraw [fill=red!8,draw=none]
	(2.5,-3) .. controls (2,0) and (1,-1)
	.. (0,0).. controls (-1,1) and (-0.9,2) .. (-1,4) -- (4,4)  -- (4,-3);
	\draw[red!60!black,very thick] (-1,4)  .. controls (-0.9,2)and (-1,1)
	.. (0,0).. controls (1,-1) and (2,0) .. (2.5,-3) node[near end, sloped, below]
	{$F(\theta)=z$};
	
	\draw [blue,very thick,dotted] (-4,-2) ellipse (6.93 and 4.9);
	\draw [blue,very thick,dotted] (-4,-2) ellipse (8.49 and 6);

	\draw [myred, line width=0.7mm] 	(4,-1) .. controls (3.95,-0.95) and (1,-1) .. (0,0).. controls (-1,1) and (-0.95,3.95) .. (-1,4);
	\node at (1.5,2.5) {\textcolor{myred}{\textbf{second-order approx.}}};
	
	\draw[black, yshift=3.5cm,xshift=5.5cm] node {\textcolor{red!60!black}{$\Omega(z)$}}; 		
	
	\filldraw [black] (0,0) circle (2pt) node[yshift=-0.5cm,xshift=0cm]{$\theta^\star$};
	\draw [black, very thick, ->] (0,0)	-- (1,1) node[yshift=-0.3cm,xshift=0.3cm]{$n^\star$};
	\end{axis}
	\end{tikzpicture}
	\caption{Two-dimensional illustration of the second-order
		approximation of the set $\Omega(z)$ for given
		$z$. These approximations exploit properties of the
		minimizer $\theta^\star$, the normal direction
		$n^\star:=\nabla F(\theta^\star)/
		\|\nabla F(\theta^\star)\|=\nabla
		I(\theta^\star)/ \|\nabla I(\theta^\star)\|$
		and the curvature
		of $\partial\Omega(z)$ at $\theta^\star$. }\label{fig:SORM}
\end{figure}
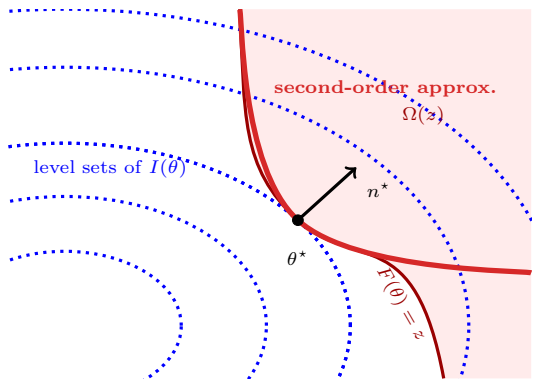

The most important step for
probability estimation using the LDT approach is to solve the
optimization problem \eqref{eq:LDT-instanton}; see also
\citep{DematteisGrafkeVandeneijnden19,dematteis2018rogue}. After
obtaining the optimizer $\theta^\star(z)$, we can integrate the
probability measure of the set bounded by the second-order Taylor
expansion \eqref{eq:FSO} to obtain an approximation of the
probability \eqref{eq:LDT-probability}, as shown in
\Cref{fig:SORM}.
For computing the prefactor $D_0(z)$ in the estimate $P(z)\approx D_0(z)
e^{-I(\theta^\star)}$, we use local derivative information of $F$ and
$I$ at $\theta^\star(z)$. The resulting expression for $D_0(z)$ is given by
\begin{align*}
D_0(z) := \dfrac{(2\pi)^{-\frac 12}}{\sqrt{2I(\theta^\star)}} \prod_{i=1}^{n-1}\left[1 -\frac{\|\nabla I(\theta^\star) \|}{\|\nabla F(\theta^\star) \|}
\lambda_i\left( \proj^\top C^{\frac 12}\nabla^2
F({\theta^\star}) C^{\frac 12} \proj \right)  \right]^{-\frac{1}{2}}.
\end{align*}
Here, for brevity, we write $\theta^\star$ instead of
$\theta^\star(z)$. The operator $\proj$ is the projection onto the
orthogonal space of the gradient $\nabla
F(\theta^\star)$. Moreover, $\lambda_i(\cdot)$ represents the $i$th
eigenvalue of a matrix. This
estimation is equivalent to the Second Order Reliability Method (SORM) in
engineering \citep{rackwitz2001reliability}. However, we
use an formulation lending itself to higher dimensions, since it only
requires the application of the second derivative matrix
$\nabla^2F(\theta^*)$ to vectors, rather than building this matrix
explicitly. We use finite differences of gradients to approximate
these Hessian-vector products. Such gradients are computed using
adjoints, as summarized in \cref{sec:opt}.
The estimation then uses randomized linear algebra methods to
compute the dominating eigenvalues efficiently
\citep{HalkoMartinssonTropp11}; details can be found in
\citep{TongVandeneijndenStadler21}.

\section{Earthquake-induced tsunamis}
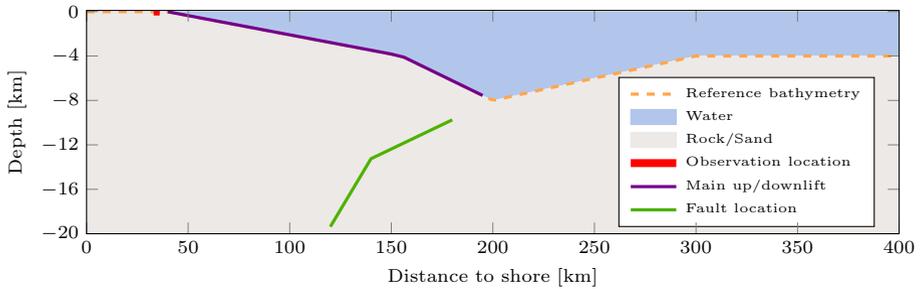
\begin{figure*}%
	\footnotesize
	\centering
	\begin{tikzpicture}[]
	\begin{axis}[compat=1.10, width=0.9\textwidth, height=0.25\textwidth, scale only axis,
	axis on top,
	xmin=0, xmax=400, ymin=-20,ymax=.1,
	xlabel={Distance to shore [km]},
	ylabel={Depth [km]},
	y tick label style={
		/pgf/number format/.cd,
		scaled y ticks = false,
		set thousands separator={,},
		fixed,
	},
	ytick = {0,-4,-8,-12, -16, -20},
	legend style={font=\tiny,nodes=right}, 
	legend pos=south east]
	\path[name path=sealevel] (axis cs: 0,0) -- (axis cs: 400,0);
	\addplot[name path=bathymetry,,orange!70,very
	thick,domain=0:400,dashed, samples=200] {1*(-0.02*(x<=40)+(-0.02-3.98/115*(x-40))*(x>40)*(x<=155)+
		(-4-4/45*(x-155))*(x>155)*(x<=200)+
		(-8+4/100*(x-200))*(x>200)*(x<=300)-
		4*(x>300))};	
	\addlegendentry{Reference bathymetry}
	\addplot[blue!75!green!30!white] fill between[ 
	of = sealevel and bathymetry, 
	soft clip={domain=0:400},
	];	\addlegendentry{Water}
	\path[name path=bottom] (axis cs: 0,-20) -- (axis cs: 400,-20);
	\addplot[orange!30!black!10!white] fill between[ 
	of =  bathymetry and bottom, 
	soft clip={domain=0:400},
	];	\addlegendentry{Rock/Sand}
	\addplot[red,line width=3pt,domain=33:36]
	{0};\addlegendentry{Observation location}
	\addplot[purple!60!blue,very thick,domain=40:195] {1*(-0.02*(x<=40)+(-0.02-3.98/115*(x-40))*(x>40)*(x<=155)+
		(-4-4/45*(x-155))*(x>155)*(x<=200)+
		(-8+4/100*(x-200))*(x>200)*(x<=300)-
		4*(x>300))};		\addlegendentry{Main up/downlift}
	\addplot [color=green!70!red, mark=none, mark size=2.5pt, very thick]
	table[x expr=\thisrowno{0}*0.001,y expr=\thisrowno{1}*0.001]
	{\data/fault.txt};\addlegendentry{Fault location}
	\end{axis}
	\end{tikzpicture}
	\caption{Problem setup inspired by Tohoku-Oki 2011
		earthquake/tsunami. Bathymetry changes (area in purple) are
		modeled as resulting from 20 randomly slipping patches in
		the fault region (in green, with end points $(120\textrm{km},-19.64\textrm{km})$ and
		$(180\textrm{km},-9.75\textrm{km})$), using the Okada model.
		The tsunami event QoI is the average wave height
		in the interval [$34$km,$35$km] close to shore (shaded in red),
		where the water depth at rest is $20$m. 
	} \label{fig:tohoprobset}
\end{figure*}

\begin{figure*}
		\footnotesize
	\centering
	\begin{tikzpicture}
	\begin{axis}[scale only axis, axis on top,
	xlabel={Horizontal location [km]},
	ylabel={Wave height [m]}, ylabel near ticks, yticklabel pos=left,
	ymax = 6.5, ymin = -3.5, xmin=30, xmax=200, width=0.9\textwidth, height= 0.15\textwidth, legend style = {font=\tiny},
	legend columns=6, 
	  ]
   \draw[very thick, fill=red, fill opacity=0.12, draw=none] (axis cs:34,-3.5) rectangle (axis cs:35,6.5);
	\addplot[smooth, color=\blueone!80, mark=none, mark size=2pt, line width=1, name path = wave5, each nth point={50}]
	table[x expr=\thisrowno{0}*0.001,y=wave] {\data/opt_T200_wv9_z4_gamma0.03.txt};
	\addplot[smooth, color=\bluetwo!80, mark=none, mark size=2pt, line width=1, name path = wave2, skip coords between index={1600}{3403}, each nth point={50}]
	table[x expr=\thisrowno{0}*0.001,y=wave] {\data/opt_T500_wv9_z4_gamma0.03.txt};
	\addplot[smooth, color=\bluethree!80, mark=none, mark size=2pt, line width=1, name path = wave3, skip coords between index={820}{3403}, each nth point={10}]
	table[x expr=\thisrowno{0}*0.001,y=wave] {\data/opt_T1000_wv9_z4_gamma0.03.txt};
	\addplot[smooth, color=\bluefour!80, mark=none, mark size=2pt, line width=1, name path = wave, skip coords between index={820}{3403}, each nth point={3}]
	table[x expr=\thisrowno{0}*0.001,y=wave] {\data/opt_max_wv9_z4_gamma0.03.txt};
	\path[name path=bottom] (axis cs: 30,-3.5) -- (axis cs: 200,-3.5);     %
	\addplot[cyan!50!blue!8] fill between[of=wave and bottom,soft clip={domain=30:200}];
	\addplot[cyan!50!blue!8] fill between[of=wave2 and bottom,soft clip={domain=30:200}];
	\addplot[cyan!50!blue!8] fill between[of=wave3 and bottom,soft clip={domain=30:200}];
	\addplot[cyan!50!blue!8] fill between[of=wave5 and bottom,soft clip={domain=30:200}];
        \node at (0.5\textwidth,0.085\textwidth)
              {\textcolor{\blueone}{$t=200$s}};
        \node at (0.28\textwidth,0.09\textwidth)
              {\textcolor{\bluetwo}{$t=500$s}};
        \node at (0.09\textwidth,0.12\textwidth)
              {\textcolor{\bluethree}{$t=1000$s}};
        \node at (0.025\textwidth,0.135\textwidth)
              {\textcolor{\bluefour}{$t_{\max}$}};
	\draw[->, very thick, -stealth, color=black!70!white]
	(axis cs:88,-2.5) --
	(axis cs: 65,-2.5) node[above, midway] {towards shore};
	\draw[->, very thick, -stealth, color=black!70!white]
	(axis cs: 175,-2.5) --
	(axis cs: 190,-2.5) node[above, midway] {towards open sea};
	\end{axis}
	\end{tikzpicture}
	\caption{Snapshots of tsunami waves generated by the optimizer
          with $z=4$ (dashed line in \Cref{fig:opt}B) at different
          times $t$. The seafloor deformation generates waves in both
          directions, but we focus on the waves traveling towards the
          shore on the left. The region where we measure average
          tsunami height at time $t_{\max}$ is shaded in red. Note
          that the tsunami wave is compressed as it travels towards
          shore, its height increases and its leading wave edge
          steepens.}
        \label{fig:snapshots}
\end{figure*}
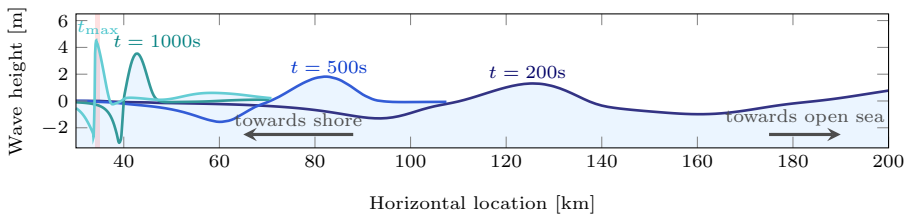

Next, we present the tsunami model. We describe how we
model the distribution of uncertain fault slips corresponding to the
random parameter $\theta$ and the forward model $F$, which involves
the solution of the shallow water equation. We use a model setup
sketched in \Cref{fig:tohoprobset}, which is inspired by the 2011 Tohoku-Oki earthquake and tsunami
\citep{FujiwaraKodairaNoEtAl11, dao2007tsunami}.
The geometry
represents a two-dimensional slice with a bathymetry that models the
continental shelf and the pacific ocean to the east of Japan.
The fault location and dip angle are taken from
\citep{zhan2012anomalously}. We generate random slips as discussed in the next section.
In \Cref{fig:snapshots}, we show
snapshots of tsunami waves traveling towards shore, which is located
on the left side of the domain.

\subsection{Modeling bathymetry change as random field} \label{sec:B}
We model earthquake-induced bathymetry change as a random variable to
account for its uncertainty. More precisely, we model
fault slips along patches in the subduction fault using a multivariate
log-normal distribution. Random fault slips are propagated to the
seafloor bathymetry change using the Okada model \citep{Okada85},
which assumes a linear elastic crust.
The use of a log-normal distribution ensures that fault slips have a
uniform sign as is realistic when stress in the overriding plate is
released through slip along the fault. We assume that slip at the
$i$th patch has the form
$s_i = \exp(\theta_i)$, where $\theta_i$ is a component of a
multivariate Gaussian distribution for the random vector $\theta$.
The choice of a mean $\hat{\mu}$ and covariance $\hat{C}$ for the
log-normal slip vector $s$ corresponds to slip events leading by
earthquakes around a certain Gutenberg-Richter moment magnitude
$M_w$. 
The slips $s=\exp(\theta)$ follow
log-normal distributions, where $\theta$ is a multivariate Gaussian
random parameter with mean $\mu$ and covariance matrix $C$.  The mean
$\mu$ and covariance matrix $C$ for $\theta$ are given in terms of
$\hat\mu$ and $\hat C$ as follows:
\begin{equation}\label{eq:GtoLN}
\begin{aligned}
C_{ij} &= \log (\hat{C}_{ij}/\hat{\mu}_i\hat{\mu}_j +1),\\
\mu_i & = \log(\hat{\mu}_i) -\frac{1}{2}C_{ii}.
\end{aligned}
\end{equation}
We next show how we choose the mean $\hat{\mu}$ and covariance $\hat
C$ such that samples of the distribution correspond to earthquaked
with a certain magnitude.
We use four pairs of $(\hat\mu,\hat C)$ and compute a weighted sum of multivariate
log-normal distributions such that the corresponding earthquakes
follow the Gutenberg-Richter scale.
The mean $\hat{\mu}$ is defined as the multiple of the taper proposed in \citep{gao2018defining}, 
\begin{equation}\label{eq:slip-mean}
\hat\mu_i := S_{M_w} \tau((x_{\text{t}} -x_i)/(x_{\text{t}} - x_{\text{b}})),
\end{equation}
where $S_{M_w}$ is a constant multiple determined by the
moment magnitude discussed later. The value $x_i$ is the horizontal location of
the center in the $i$th sub-fault, and $x_{\text{t}} $ and $x_{\text{b}}$
are the horizontal location of the top and the bottom of the fault,
respectively. The function $\tau$ is the taper from
\citep{gao2018defining}:
\begin{equation}
\tau(x') = \delta(x')(1+\sin(\pi\delta(x')^b)),
\end{equation}
\begin{equation}
\delta(x')=\left\lbrace
\begin{aligned}
&\frac{6}{q^3}(x')^2\left(\frac{q}{2}-\frac{x'}{3}\right), && 0\leq x'\leq q,\\
&\frac{6}{(1-q)^3}(1-x')^2\left(\frac{1-q}{2}-\frac{1-x'}{3}\right), && q\leq x'\leq 1.
\end{aligned}
\right. 
\end{equation}
We use the values $b=0.25$, $q=0.65$. The covariance is defined similarly as in
\citep{leveque2016generating}, namely as
\begin{equation}\label{eq:slip-cov}
\hat{C}_{ij}= (3/4)^2 \hat{\mu}_i \hat{\mu}_j
\exp(-\lvert x_i-x_j\rvert / (0.4\lvert x_{\text{t}} - x_{\text{b}}\rvert)). 
\end{equation}
The only thing left is to choose $S_{M_w}$ such that $\hat{\mu}$
produce to an earthquake with moment magnitude $M_w$, by solving the
relation
\begin{equation}
M_w = \frac{2}{3} (\log_{10}M_0-9.05),
\end{equation}
where the seismic moment $M_0$ = $A \times $ (average slip) $\times$
(rigidity). Following \citep{murotani2008scaling}, the area $A$ is defined as
\begin{equation}
\label{eq:area}
A=1.48\cdot 10^{-10+9.05\cdot 2/3+6+M_w} [m^2].
\end{equation}
The average overall
slip is computed as the average of \eqref{eq:slip-mean}. The rigidity is
$35\times 10^9 N/m^2$, as suggested in \citep{hashima2016coseismic}.

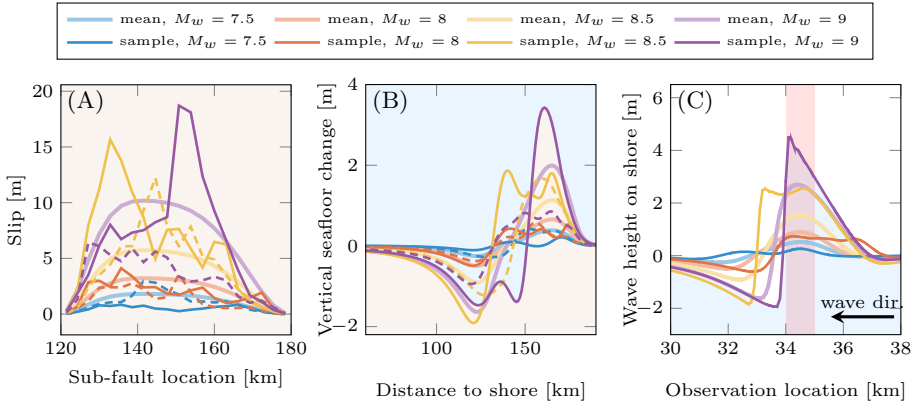
\begin{figure*}
		\footnotesize
	\centering
	\begin{tikzpicture}[]
	\begin{groupplot}[group style = {rows=1, columns=3, horizontal
              sep = 28pt, vertical sep=0pt,}, width = 0.255\textwidth, height =
             0.28\textwidth,
             axis on top,
             ]
	\nextgroupplot[compat=1.3, scale only axis,
	xlabel={Sub-fault location [km]},
	ylabel={Slip [m]},
	legend style = {font=\tiny,nodes=right, legend to name=grouplegend},
	legend columns=4, 
	xmin=120,xmax=180, ylabel shift=-1ex, 
	axis background/.style={fill=brown!90!black!8}]
	\addplot[color=\cOne!40, line width=1.5, mark=none, mark size=1.5pt,]
	table[x expr=\thisrowno{0}*0.001,y=mean] {\data/sample_slip7.5.txt};
	\addlegendentry{mean, $M_w=7.5$}
	\addplot[color=\cTwo!40, line width=1.5, mark=none, mark size=1.5pt,]
	table[x expr=\thisrowno{0}*0.001,y=mean] {\data/sample_slip8.txt};
	\addlegendentry{mean, $M_w=8$}
	\addplot[color=\cThree!40,  line width=1.5, mark=none, mark size=1.5pt,]
	table[x expr=\thisrowno{0}*0.001,y=mean] {\data/sample_slip8.5.txt};
	\addlegendentry{mean, $M_w=8.5$}
	\addplot[color=\cFour!40,  line width=1.5, mark=none, mark size=1.5pt,]
	table[x expr=\thisrowno{0}*0.001,y=mean] {\data/sample_slip9.txt};
	\addlegendentry{mean, $M_w=9$}
	\addplot[color=\cOne!80, line width=1]
	table[x expr=\thisrowno{0}*0.001,y=A] {\data/sample_slip7.5.txt};
	\addlegendentry{sample, $M_w=7.5$}
	\addplot[color=\cOne!80, line width=1, forget plot, densely dashed]
	table[x expr=\thisrowno{0}*0.001,y=B] {\data/sample_slip7.5.txt};
		\addplot[color=\cTwo!80, line width=1]
	table[x expr=\thisrowno{0}*0.001,y=A] {\data/sample_slip8.txt};
	\addlegendentry{sample, $M_w=8$}
	\addplot[color=\cTwo!80, line width=1,forget plot, densely dashed]
	table[x expr=\thisrowno{0}*0.001,y=B] {\data/sample_slip8.txt};
		\addplot[color=\cThree!80, line width=1]
	table[x expr=\thisrowno{0}*0.001,y=A] {\data/sample_slip8.5.txt};
	\addlegendentry{sample, $M_w=8.5$}
	\addplot[color=\cThree!80, line width=1, forget plot, densely dashed]
	table[x expr=\thisrowno{0}*0.001,y=B] {\data/sample_slip8.5.txt};
		\addplot[color=\cFour!80, line width=1]
	table[x expr=\thisrowno{0}*0.001,y=A] {\data/sample_slip9.txt};
	\addlegendentry{sample, $M_w=9$}
	\addplot[color=\cFour!80, line width=1, forget plot,   densely dashed]
	table[x expr=\thisrowno{0}*0.001,y=B] {\data/sample_slip9.txt};
	\nextgroupplot[scale only axis,
	xlabel={Distance to shore [km]},
	ylabel={Vertical seafloor change [m]}, ,ylabel near ticks, yticklabel pos=left,
	xmin=60, xmax=190, ylabel shift=-2.5ex, ymax = 4, ymin = -2.2
	]
	\addplot [color=\cOne!40, mark=none, mark size=2pt, line width=1.5]
	table[x expr=\thisrowno{0}*0.001,y=mean] {\data/sample_deformation7.5.txt};
	\addplot [color=\cTwo!40, mark=none, mark size=2pt, line width=1.5]
	table[x expr=\thisrowno{0}*0.001,y=mean] {\data/sample_deformation8.txt};
	\addplot [color=\cThree!40, mark=none, mark size=2pt, line width=1.5]
	table[x expr=\thisrowno{0}*0.001,y=mean] {\data/sample_deformation8.5.txt};
	\addplot [color=\cFour!40, mark=none, mark size=2pt, line width=1.5]
	table[x expr=\thisrowno{0}*0.001,y=mean] {\data/sample_deformation9.txt};
	\addplot [color=\cOne!80, mark=none, mark size=2pt, line width=1]
	table[x expr=\thisrowno{0}*0.001,y=A] {\data/sample_deformation7.5.txt};
	\addplot [color=\cOne!80, mark=none, mark size=2pt, line width=1,  densely dashed]
	table[x expr=\thisrowno{0}*0.001,y=B] {\data/sample_deformation7.5.txt};
	\addplot [color=\cTwo!80, mark=none, mark size=2pt, line width=1]
table[x expr=\thisrowno{0}*0.001,y=A] {\data/sample_deformation8.txt};
	\addplot [color=\cTwo!80, mark=none, mark size=2pt, line width=1,  densely dashed]
table[x expr=\thisrowno{0}*0.001,y=B] {\data/sample_deformation8.txt};
	\addplot [color=\cThree!80, mark=none, mark size=2pt, line width=1]
table[x expr=\thisrowno{0}*0.001,y=A] {\data/sample_deformation8.5.txt};
	\addplot [color=\cThree!80, mark=none, mark size=2pt, line width=1, densely dashed]
table[x expr=\thisrowno{0}*0.001,y=B] {\data/sample_deformation8.5.txt};
	\addplot[color=\cFour!80, mark=none, mark size=2pt, line width=1, name path = deformation, each nth point={10}]
table[x expr=\thisrowno{0}*0.001,y=A] {\data/sample_deformation9.txt};
\addplot [color=\cFour!80, mark=none, mark size=2pt, line width=1, densely dashed]
table[x expr=\thisrowno{0}*0.001,y=B] {\data/sample_deformation9.txt};
\path[name path=earth] (axis cs: 60,-2.2) -- (axis cs: 190,-2.2);
\path[name path=sea] (axis cs: 60,4) -- (axis cs: 190,4);
\addplot[cyan!50!blue!8] fill between[of=sea and deformation,soft clip={domain=60:190}]; %
\addplot[brown!90!black!8] fill between[of=deformation and earth, soft clip={domain=60:190}]; %
\nextgroupplot[scale only axis,
	xlabel={Observation location [km]},
ylabel={Wave height on shore [m]}, ylabel near ticks, yticklabel pos=left,
ymax = 6.5, ymin = -3, xmin=30, xmax=38, ylabel shift=-2.5ex
]
   \draw[very thick, fill=red, fill opacity=0.12, draw=none] (axis cs:34,-3) rectangle (axis cs:35,6.5);
	\addplot[color=\cOne!40, line width=1.5, mark=none, mark size=1.5pt,]
table[x expr=\thisrowno{0}*0.001,y=wave] {\data/mean_wv7.5_gamma0.03.txt};
\addplot[color=\cTwo!40, line width=1.5, mark=none, mark size=1.5pt,]
table[x expr=\thisrowno{0}*0.001,y=wave] {\data/mean_wv8_gamma0.03.txt};
\addplot[color=\cThree!40,  line width=1.5, mark=none, mark size=1.5pt,]
table[x expr=\thisrowno{0}*0.001,y=wave] {\data/mean_wv8.5_gamma0.03.txt};
\addplot[color=\cFour!40,  line width=1.5, mark=none, mark size=1.5pt,]
table[x expr=\thisrowno{0}*0.001,y=wave] {\data/mean_wv9_gamma0.03.txt};
	\addplot[color=\cOne!80, line width=1, mark=none, mark size=1.5pt,]
	table[x expr=\thisrowno{0}*0.001,y=wave] {\data/sample_wv7.5_gamma0.03.txt};
	\addplot[color=\cTwo!80, line width=1, mark=none, mark size=1.5pt,]
	table[x expr=\thisrowno{0}*0.001,y=wave] {\data/sample_wv8_gamma0.03.txt};
	\addplot[color=\cThree!80,  line width=1, mark=none, mark size=1.5pt,]
	table[x expr=\thisrowno{0}*0.001,y=wave] {\data/sample_wv8.5_gamma0.03.txt};
	\addplot[color=\cFour!80,  line width=1, mark=none, mark size=1.5pt, name path = wave]
	table[x expr=\thisrowno{0}*0.001,y=wave] {\data/sample_wv9_gamma0.03.txt};
	\path[name path=bottom] (axis cs: 38,-3) -- (axis cs: 30,-3);     %
	\addplot[cyan!50!blue!8] fill between[of=wave and bottom,soft clip={domain=30:38}]; %
	\end{groupplot}
	\node[black] at ($(group c1r1) + (0.32\textwidth,0.2\textwidth)$) {\pgfplotslegendfromname{grouplegend}}; 
	\node at ($(group c1r1) +(-0.1\textwidth,0.12\textwidth)$) {\large \textcolor{black}{\small (A)}};
\node at ($(group c2r1) +(-0.1\textwidth,0.12\textwidth)$) {\large \textcolor{black}{\small (B)}};
\node at ($(group c3r1) +(-0.1\textwidth,0.12\textwidth)$) {\large \textcolor{black}{\small (C)}};
\draw[->, very thick, -stealth] ($(group c3r1) +(0.12\textwidth,-0.12\textwidth)$) -- ($(group c3r1) +(0.05\textwidth,-0.12\textwidth)$) node[above, midway] {wave dir.};
	\end{tikzpicture}
	\caption{Shown in (A) are random slip patterns (solid and
          dashed lines) from the log-normal distributions whose means
          (faded solid lines) generate earthquake with moment
          magnitudes $M_w = 7.5, 8, 8.5, 9$. Different colors indicate
          slips originating from distributions with different means.
          Shown in (B) are the vertical seafloor deformation induced
          by the slip samples and means shown in (A). These seafloor
          deformations are computed using the Okada model.  Shown in
          (C) are the tsunami waves close to shore generated by the
          solid line samples and the mean seafloor deformations in
          (B).  The tsunami waves are shown at the time when they
          maximize the objective \eqref{eq:G}, i.e., their average
          wave height is maximal in the measurement interval (red
          shaded area).
	}\label{fig:sample}
\end{figure*}

The framework summarized in \cref{sec:LDT} uses the
multivariate Gaussian variable $\theta$ underlying the log-normal
variable $s$. 
We use four mean slips and covariances for
earthquakes of moment magnitude $M_w=7.5,8,8.5,9$. These mean slips,
as well as random draws from the four log-normal distributions are
shown in \Cref{fig:sample}A. Note that this distribution assumes
that the faults are of buried type, i.e., there is no slip close to
the seafloor. By changing the means, other scenarios such as
trench-breaking earthquakes can be modeled
\citep{gao2018defining}. The bathymetry changes induced by mean and
random fault slips are shown in \Cref{fig:sample}B.

Note that the proposed method for estimating the probability of the
tsunami height on shore is not limited to the slip distributions used
in this section (i.e., log-normal distributions for slips means and
the exponential covariance function \eqref{eq:slip-cov} as suggested
in \citep{leveque2016generating}). The approach is generic and can be
modified to other settings that have been proposed in the
literature. For example, it is straightforward to change the
exponential dependency of the covariance \eqref{eq:slip-cov} from the
distance $\lvert x_i-x_j\rvert $ to the von Karman
distribution \citep{mai2002spatial, crempien2020effects}. This will
only impact the definitions for mean and covariance
in \eqref{eq:GtoLN}, 
but all other steps remain unchanged. Also
heavy-tailed distributions built by discrete Fourier transformations
and power law dependency of the
spectrum \citep{lavallee2003stochastic, lavallee2006stochastic} could be
used. This would require to incorporate
the discrete Fourier functions into the map $F$, similar to the way we
incorporate the exponential transformation into $F$ to obtain
log-normal distributions.

\subsection{Governing nonlinear shallow water equations}\label{sec:SWE}
We model tsunami waves using the one-dimensional shallow water
equations, describing the conservation of mass and momentum for points
$x$ in space and times $t\in[0,T]$, $T>0$.  These nonlinear
hyperbolic equations, written in terms of the water height $h$ and the
momentum variable $v$ are:
\begin{subequations}\label{eq:shallow}
\begin{eqnarray}
    h_t+v_x&=0,\label{eq:shallow1}\\
    v_t+\left(\frac{v^2}{h}+\frac{1}{2}gh^2\right)_{\!\!x}+ghB_x&=0.\label{eq:shallow2}
\end{eqnarray}
\end{subequations}
Here, $v:=hu$ with $u$ being the velocity, $g$ is the gravitational
constant, and $B$ the earthquake-induced bathymetry change, which
enters through its spatial derivative $B_x$.  The sea
is assumed to be in rest at initial time $t=0$ for the given
bathymetry $B_0$, i.e., $v(x,0)=0$ and $h(x,0)=-B_0(x)$ for all $x$.
Together with these initial conditions, we assume suitable boundary
conditions that are sufficiently far away from the observation
interval and the source such they do not interact with the solution in
the considered time interval.
The shallow water model equations \eqref{eq:shallow} do not take into
account a horizontal bathymetry change. Any non-zero vertical
bathymetry change resulting from a fault slip lifts the water column
and thus leads to waves traveling in both directions, towards and away
from shore; see \Cref{fig:snapshots}. Next, we describe our measure of tsunami size on shore.

\subsection{Parameter-to-QoI map and combining Gaussians}\label{sec:setup-summary}
We use the average tsunami wave height in a small interval $[c,d]$ close to shore to measure the tsunami size. This interval is indicated in red in
\Cref{fig:snapshots} and in \Cref{fig:tohoprobset}. Since we
cannot exactly predict how long after the bathymetry change the
largest tsunami wave reaches shore, we approximate the maximum
average wave height using a smoothing parameter $\gamma>0$,
resulting in
\begin{equation}
\label{eq:Greg}
G_\gamma(B):=
\gamma\log\left[ 
\dfrac{1}{T}\int_{0}^{T}\exp \left(\dfrac{1}{\gamma}\fint_c^d(h+B_0)dx\right) dt\right].
\end{equation}
Here, $h$ is considered a function of $B$ through the solution of the
shallow water equations \eqref{eq:shallow}. Moreover, $h+B_0$ is the
water height above the resting state, and $\fint$ denotes the average
integral over $[c,d]$. It can be shown that
\begin{equation}\label{eq:G}
\lim_{\gamma\rightarrow 0} G_\gamma(B) = \max\limits_{t\in[0,T_F]} \fint_c^d [h(x,t)+B_0(x)]dx.
\end{equation}
i.e., for large $\gamma$, \eqref{eq:Greg} approximately measures the
largest average wave height in $[c,d]$ over all times in $[0,T]$.

We apply the approach summarized in \cref{sec:LDT} individually to the
multivariate Gaussians underlying the sum of log-normal distributions
used to model the distribution of fault slips. The parameter-to-QoI
map $F$ includes the transformation from Gaussian to log-normal
parameters, the map from slip patches to bathymetry change
$B$ governed by the Okada model, and the solution of the shallow water
equations to map the bathymetry change to the average wave height
close to shore \eqref{eq:Greg}, i.e., $F$ is defined as:
\begin{equation}\label{eq:Ftsunami}
F:\theta\mapsto \exp(\theta) \mapsto B \mapsto G_\gamma(B).
\end{equation}
Components of this map are illustrated in \Cref{fig:sample}.
Note that all samples from the bathymetry change
distribution are negative on the left, i.e., closer to
shore, and positive on the right. This is a consequence of only
using positive slips reflecting the mechanisms behind plate
subduction, i.e., the overriding plate releases stress during fault
slip events. Due to the structure of these random slip-induced bathymetry changes, the
trough of a tsunami wave reaches the shore first, followed by the wave
crest as seen in \Cref{fig:sample}C.

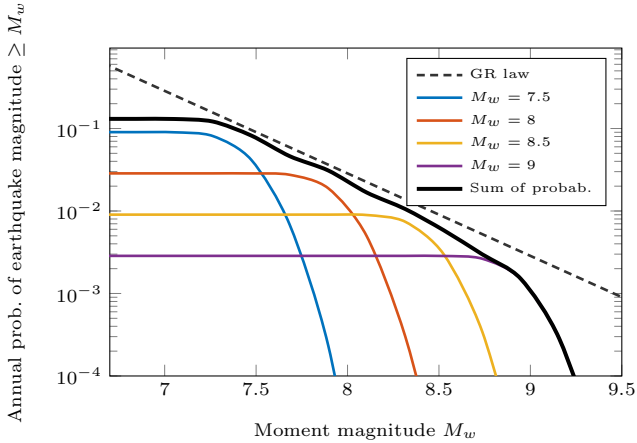
\begin{figure}
	\footnotesize
	\centering
	\begin{tikzpicture}
	\begin{semilogyaxis}[
	height=0.5\textwidth, 
	width = 0.7\textwidth, 
	legend cell align={left},
	legend pos=north east,
	legend style = {font=\tiny},
	xlabel={Moment magnitude $M_w$}, ylabel={Annual prob.\ of earthquake magnitude
		$\geq M_w$ },
	xmin=6.7, xmax=9.5, ymin=1e-4
	]
	\addplot[color=black!80, densely dashed,  line width=1, mark=none, mark size=1.5pt] table [x=x,y=GR] {\data/MwDistribution.txt};\addlegendentry{GR law}
	\addplot[smooth, color=\cOne,  line width=1, mark=none, mark size=1.5pt]  table [x=x,y=prob] {\data/MwDistribution_Mw7.5.txt}; \addlegendentry{$M_w=7.5$}
	\addplot[smooth, color=\cTwo,  line width=1, mark=none, mark size=1.5pt]  table [x=x,y=prob] {\data/MwDistribution_Mw8.txt}; \addlegendentry{$M_w=8$}
	\addplot[smooth, color=\cThree,  line width=1, mark=none, mark size=1.5pt]  table [x=x,y=prob] {\data/MwDistribution_Mw8.5.txt}; \addlegendentry{$M_w=8.5$}
	\addplot[smooth, color=\cFour,  line width=1, mark=none, mark size=1.5pt]  table [x=x,y=prob] {\data/MwDistribution_Mw9.txt}; \addlegendentry{$M_w=9$}
	\addplot[smooth, color=black,  line width=1.5, mark=none, mark
	size=1.5pt]  table [x=x,y=mix] {\data/MwDistribution.txt}; \addlegendentry{Sum of probab.}
	\end{semilogyaxis}
	\end{tikzpicture}
	\caption{Annual probabilities of earthquakes magnitude
		larger than $M_w$ caused by samples generated from $\pi_{M_w}$
		whose means correspond to moment magnitude
		$M_w= 7.5, 8, 8.5, 9$.  The sum of the four probability
		distributions provides the annual probabilities of observing
		an earthquake with magnitude from 7 to 9. We compare it with the probability
		distribution $10^{6.456-M}$ from the Gutenberg-Richter law. It can be seen that our model fits the
		distribution of earthquakes for magnitude from 7 to 9.
	}
	\label{fig:Mw}
\end{figure}

The distribution is the sum of four log normals with different means,
which are combined such that the resulting distribution of earthquakes
follows the Gutenberg-Richert (GR) law for earthquakes with moment
magnitude ranging from 7 to 9. 
The previous sections detail the probability distribution for $\theta$
with a fixed mean corresponding to moment magnitude $M_w$. We denote
this multivariate Gaussian distribution as $\pi_{M_w}$.
For this setup, we apply the methods discussed in \cref{sec:LDT}
to estimate $\PP_{\pi_{M_w}}(F(\theta)\geq z)$, for the probability of
observing an average wave height of $z$ or higher from random slips
from the distribution with moment magnitude $M_w$ mean. To obtain the
annual assessment of this probability $P_{\text{an}}(z)$, i.e., the
annual probability of wave higher than $z$, we use a weighted sum
similar to \citep{williamson2020source}
\begin{equation}
P_{\text{an}}(z) = \sum_{M_w \in\{7.5, 8, 8.5, 9\} } w_{M_w} \PP_{\pi_{M_w}}(F(\theta)\geq z),
\end{equation}
where the weight $w_{M_w}=10^{6.456-M_w}$ is the annual probability of
a moment magnitude $M_w$ earthquake following the Gutenberg-Richter
(GR) law.  The Gutenberg-Richter (GR) law describes the return period
of certain magnitude earthquakes. We choose 350 years as the return
period for the Tohoku area from the study in \citep{kagan2013tohoku},
so the annual probability of occurring earthquakes with moment
magnitude larger than $M_w$ is $10^{6.456-M_w}$. Using these
weights and the distribution $\pi_{M_w}$, we compute the annual
probability for earthquakes with magnitude larger than $M_w$ when
samples are from $\pi_{M_w}$. Their sum fits the annual probability
curve $10^{6.456-M_w}$ from the GR law, as shown in \Cref{fig:Mw}.

\section{Numerical and optimization methods}\label{sec:opt}

Our approach for probability estimation requires to solve a sequence
of optimization problems of the form \eqref{eq:LDT-instanton}
with the parameter-to-event map $F$ specified above. Since $F$
involves solution of the shallow water equations for given $B$ (and
thus $\theta$), this is a PDE-constrained optimization problem
\citep{borzi2011computational,HinzePinnauUlbrichEtAl09,Reyes15}. For
solving the one-dimensional shallow water equations, we use the
discontinuous Galerkin finite element method (DG-FEM) with linear
interpolating polynomials and a global Lax-Friedrichs flux to
discretize the equations in space, and the strong stability-preserving
second-order Runge-Kutta (SSP-RK2) method to discretize the equations
in time \citep{hesthaven2007nodal}.  We use adjoint methods to
efficiently compute gradients for this optimization problem, and a
descent algorithm that approximates second-order derivative
information using the BFGS method \citep{NocedalWright06}. After
finding the optimizer, we use finite differences of gradients to
approximate the application of Hessians to vectors as required to find
the dominating curvature directions; for details we refer to
\citep{TongVandeneijndenStadler21}.
Numerical values used in the discretization the problem is summarized in \Cref{tab:number}.

\begin{table}[h]
	{
		\caption{Numerical values used for problem setup
			and discretization.}\label{tab:number}
		\begin{center}
			\begin{tabular}{l|l} \hline
				Number of elements using in DG-FEM &
				5000 \\ \hline
				Final Time $T$ & 1800 \\ \hline
				Time step size & 0.0714 \\  \hline
				Smoothing parameter $\gamma$ & 0.03 \\
				\hline
			\end{tabular}
		\end{center}
	}
\end{table}

As we have seen, solutions of nonlinear hyperbolic equations such as
the shallow water equations can develop steep slopes or shocks, which
play an important role for the dynamics of the system. These phenomena
are challenging for numerical simulations. To prevent infinite slopes,
in our simulations we use artificial viscosity, which decreases upon
mesh refinement, thus retaining important aspects of the dynamics,
details can be found in \citep{TongVandeneijndenStadler21}.

\section{Results}

\begin{figure}
		\footnotesize
	\centering
	\begin{tikzpicture}
	\begin{semilogyaxis}[
	height=0.37*42pc, width = 0.45*42pc,
	legend cell align={left},
	legend pos=north east,
	legend style = {font=\tiny}, legend columns=2, 
xlabel={Tsunami height threshold $z$ [m]}, ylabel={Annual prob.\ of
  tsunami height $\ge z$}, xmin=0.5, xmax=10.2, ymin=1e-8
	]
	\addplot[color=black!50,  line width=1.5, mark=none, mark
          size=1.5pt, densely dashed]  table [x=z,y=prob]
                {\data/probMC_mix_gamma0.03.txt};\addlegendentry{MC sum}
	\addplot[color=black,  line width=1, mark=*, mark size=1.5pt,
          ]  table [x=z,y=prob]
                {\data/prob_mix_gamma0.03.txt};\addlegendentry{LDT sum} 
	\addplot[color=\cOne!40,  line width=1, mark=*, mark
          size=1.5pt]  table [x=z,y=prob]
                {\data/prob_annual7.5_gamma0.03.txt};
                \addlegendentry{$M_w=7.5$}
	\addplot[color=\cTwo!40,  line width=1, mark=*, mark
          size=1.5pt]  table [x=z,y=prob]
                {\data/prob_annual8_gamma0.03.txt};
                \addlegendentry{$M_w=8$}
	\addplot[color=\cThree!40,  line width=1, mark=*, mark
          size=1.5pt]  table [x=z,y=prob]
                {\data/prob_annual8.5_gamma0.03.txt};
                \addlegendentry{$M_w=8.5$}
	\addplot[color=\cFour!40,  line width=1, mark=*, mark
          size=1.5pt]  table [x=z,y=prob]
                {\data/prob_annual9_gamma0.03.txt};
                \addlegendentry{$M_w=9$}
        \addplot[color=\cOne!40,  line width=1, mark=*, mark
          size=2.5pt, 
          only marks, 
          skip coords between index={1}{10},
          forget plot, name path = opt1]  table [x=z,y=prob]
                {\data/prob_annual7.5_gamma0.03.txt};
        \addplot[color=\cThree!40,  line width=1, mark=*, mark
          size=2.5pt, skip coords between index={0}{1}, skip coords
          between index={2}{10}, forget plot, only marks]  table [x=z,y=prob]
                {\data/prob_annual8.5_gamma0.03.txt};
	\addplot[color=\cFour!40,  line width=1, mark=*, mark
          size=2.5pt, forget plot, only marks, skip coords between index={0}{2},
        skip coords between index={5}{10}]  table [x=z,y=prob]
                {\data/prob_annual9_gamma0.03.txt};
	\end{semilogyaxis}
        \node at (0.019*42pc,0.236*42pc) {\textcolor{black!90}{$a$}};
        \node at (0.056*42pc,0.212*42pc) {\textcolor{black!90}{$b$}};
        \node at (0.093*42pc,0.196*42pc) {\textcolor{black!90}{$c$}};
        \node at (0.13*42pc,0.186*42pc) {\textcolor{black!90}{$d$}};
        \node at (0.167*42pc,0.171*42pc) {\textcolor{black!90}{$e$}};
	\end{tikzpicture}
	\caption{Comparison of the annual probability of
          $F(\theta)\ge z$, i.e., tsunami waves on shore of
          size $z$ or larger between Monte Carlo sampling (dashed lines) and LDT
          approximation (solid lines). Shown in color are the contribution of the
          log-normal distributions centered at magnitude 7.5, 8, 8.5 and
          9 to the overall probability as discussed in \cref{sec:setup-summary}. The points for
          $z=1,\ldots, 5$, labeled with (a)--(e), correspond to the
          optimized fault slips shown in
          \Cref{fig:opt}A and represent the dominant contribution
          to the overall probability. Monte Carlo sampling uses
          $10^4$ samples for each of the four earthquake magnitudes.
      }
	\label{fig:prob}
\end{figure}
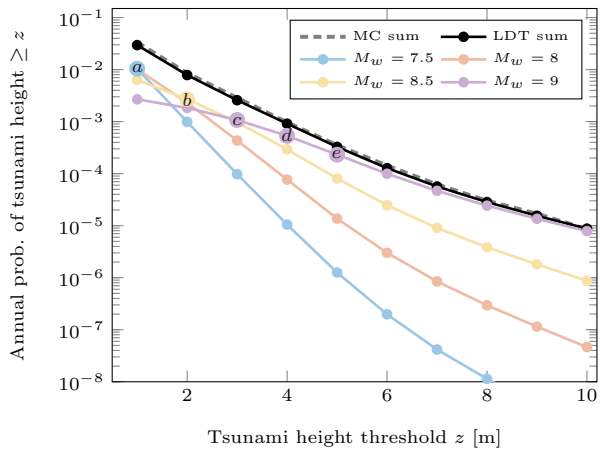

First, we compare the estimation of tsunami wave height probabilities from
the optimization-based method with the probability estimation from
Monte Carlo sampling. In \Cref{fig:prob} it can be seen that the
sampling-based and the LDT-based probability estimates are visually
indistinguishable. In this figure we also illustrate how the
log-normal distributions around the four different slip means
(corresponding to earthquakes of different magnitude) contribute to
the overall probability. Note that for large tsunami waves, the
distribution centered at magnitude $M_w=9$ event dominates the
probability.  While typically Monte Carlo requires a large number of
samples to estimate low probabilities, this effect is mitigated here
by the fact that the small probabilities are dominated by the events
in the distribution around the magnitude 9 earthquake mean.

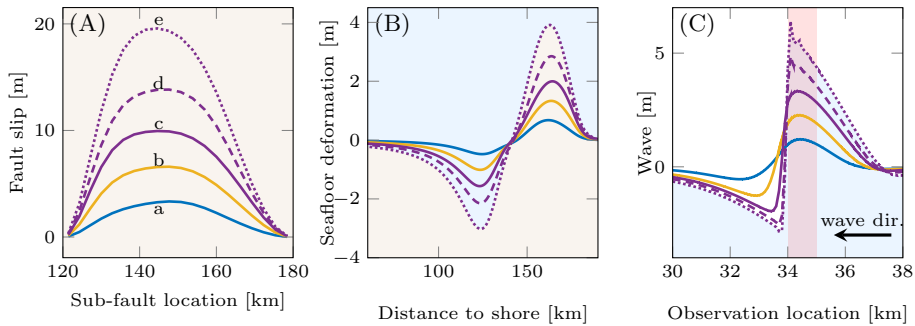
\begin{figure*}
		\footnotesize
	\centering
	\begin{tikzpicture}[]
	\begin{groupplot}[group style = {rows=1, columns=3, horizontal
              sep = 28pt, vertical sep=0pt,}, width =  0.255\textwidth, height =
             0.28\textwidth,
             axis on top,
             ]
	\nextgroupplot[compat=1.3, scale only axis,
	xlabel={Sub-fault location [km]},
	ylabel={Fault slip [m]},
	legend style = {font=\tiny,nodes=right, legend to name=grouplegend2},
	legend columns=5, 
	xmin=120,xmax=180,ylabel shift=-1ex,
	axis background/.style={fill=brown!90!black!8}]
	\addplot[color=\cOne, line width=1, mark=none, mark size=1.5pt,]
	table[x expr=\thisrowno{0}*0.001,y=slip] {\data/opt_slip7.5_z1_gamma0.03.txt} ;
	\addplot[color=\cThree, line width=1, mark=none, mark size=1.5pt,]
	table[x expr=\thisrowno{0}*0.001,y=slip] {\data/opt_slip8.5_z2_gamma0.03.txt};
	\addplot[color=\cFour, line width=1, mark=none, mark size=1.5pt,]
	table[x expr=\thisrowno{0}*0.001,y=slip] {\data/opt_slip9_z3_gamma0.03.txt};
		\addplot[densely dashed, color=\cFour, line width=1, mark=none, mark size=1.5pt,]
	table[x expr=\thisrowno{0}*0.001,y=slip] {\data/opt_slip9_z4_gamma0.03.txt};
	\addplot[densely dotted, color=\cFour, line width=1, mark=none, mark size=1.5pt,]
	table[x expr=\thisrowno{0}*0.001,y=slip] {\data/opt_slip9_z5_gamma0.03.txt};
	\node at (axis cs: 145, 2.5) {\textcolor{black}{a}};
	\node at (axis cs: 145, 7.3) {\textcolor{black}{b}};
	\node at (axis cs: 145, 10.5) {\textcolor{black}{c}};
	\node at (axis cs: 145, 14.5) {\textcolor{black}{d}};
	\node at (axis cs: 145, 20.3) {\textcolor{black}{e}};
	\nextgroupplot[scale only axis,
	xlabel={Distance to shore [km]},
	ylabel={Seafloor deformation [m]}, ,ylabel near ticks, yticklabel pos=left,
	ylabel shift=-2.5ex, xmin=60, xmax=190, ymax=4.5, ymin = -4
	]
	\addplot[color=\cOne, line width=1, mark=none, mark size=1.5pt,]
	table[x expr=\thisrowno{0}*0.001,y=deform] {\data/opt_deform7.5_z1_gamma0.03.txt};
	\addplot[color=\cThree, line width=1, mark=none, mark size=1.5pt,]
	table[x expr=\thisrowno{0}*0.001,y=deform] {\data/opt_deform8.5_z2_gamma0.03.txt};
	\addplot[color=\cFour, line width=1, mark=none, mark size=1.5pt,]
	table[x expr=\thisrowno{0}*0.001,y=deform] {\data/opt_deform9_z3_gamma0.03.txt};
	\addplot[densely dashed, color=\cFour, line width=1, mark=none, mark size=1.5pt,]
	table[x expr=\thisrowno{0}*0.001,y=deform] {\data/opt_deform9_z4_gamma0.03.txt};
	\addplot[densely dotted, color=\cFour, line width=1, mark=none, mark size=1.5pt, name path = deformation, each nth point = {10}]
	table[x expr=\thisrowno{0}*0.001,y=deform] {\data/opt_deform9_z5_gamma0.03.txt};
	\path[name path=earth] (axis cs: 60,-4) -- (axis cs: 190,-4);
	\path[name path=sea] (axis cs: 60,4.5) -- (axis cs: 190,4.5);
	\addplot[cyan!50!blue!8] fill between[of=sea and deformation,soft clip={domain=60:190}]; %
	\addplot[brown!90!black!8] fill between[of=deformation and earth, soft clip={domain=60:190}]; %
	\nextgroupplot[scale only axis,
	xlabel={Observation location [km]},
	ylabel={Wave [m]}, ylabel near ticks, yticklabel pos=left,
	ymax = 7, ymin = -4, 
	ylabel shift=-2ex,
	 xmin=30, xmax=38
	]
   \draw[very thick, fill=red, fill opacity=0.12, draw=none] (axis cs:34,-4) rectangle (axis cs:35,7);
	\addplot[color=\cOne, line width=1, mark=none, mark size=1.5pt,]
table[x expr=\thisrowno{0}*0.001,y=wave] {\data/opt_wv7.5_z1_gamma0.03.txt};
\addplot[color=\cThree, line width=1, mark=none, mark size=1.5pt,]
table[x expr=\thisrowno{0}*0.001,y=wave] {\data/opt_wv8.5_z2_gamma0.03.txt};
\addplot[color=\cFour, line width=1, mark=none, mark size=1.5pt,]
table[x expr=\thisrowno{0}*0.001,y=wave] {\data/opt_wv9_z3_gamma0.03.txt};
\addplot[densely dashed, color=\cFour, line width=1, mark=none, mark size=1.5pt, ]
table[x expr=\thisrowno{0}*0.001,y=wave] {\data/opt_wv9_z4_gamma0.03.txt};
\addplot[densely dotted, color=\cFour, line width=1, mark=none, mark size=1.5pt, name path = shock]
table[x expr=\thisrowno{0}*0.001,y=wave] {\data/opt_wv9_z5_gamma0.03.txt};
\path[name path=bottom] (axis cs: 38,-4) -- (axis cs: 30,-4);     %
\addplot[cyan!50!blue!8] fill between[of=shock and bottom,soft clip={domain=30:38}];
\draw[->, very thick, -stealth] (axis cs: 37.6, -3) -- (axis cs: 35.6, -3) node[above, midway] {wave dir.};
	\end{groupplot}
	\node[black] at ($(group c1r1) + (6cm,3.0cm)$) {\pgfplotslegendfromname{grouplegend2}}; 
	\node at ($(group c1r1) +(-0.1\textwidth,0.12\textwidth)$) {\large \textcolor{black}{\small (A)}};
\node at ($(group c2r1) +(-0.1\textwidth,0.12\textwidth)$) {\large \textcolor{black}{\small (B)}};
\node at ($(group c3r1) +(-0.1\textwidth,0.12\textwidth)$) {\large \textcolor{black}{\small (C)}};
	\end{tikzpicture}
	\caption{Shown in (A) are the slips corresponding to the
          LDT-optimizers that contributes most to the overall
          probability. These probabilities are indicated by the solid
          colored spheres in \Cref{fig:prob}. The dominant
          contributions for $z=1$ is from the log-normal distribution
          with mean $M_w=7.5$, for $z=2$ from the distribution with
          mean $M_w=8.5$ and for $z=3,4,5$ from the distribution with
          mean $M_w=9$. Shown in the middle are the corresponding
          vertical seafloor deformations.  Shown in (C) are the
          tsunami waves close to shore generated by the seafloor
          deformations in (B).  Snapshots of the waves are shown when
          \eqref{eq:Greg}, i.e., the average wave height is largest in
          the measurement interval (red shaded area).
      }\label{fig:opt}
\end{figure*}

The fault slips dominating the probabilities for $z=1,\ldots, 5$
meters, which are computed as minimizers of \eqref{eq:LDT-instanton},
are labeled in \Cref{fig:prob}.
These
correspond to the most likely slip mechanism that results in a tsunami
on shore of size $z$. These slips, the corresponding seafloor
deformation and the resulting tsunami waves on shore are shown in
\Cref{fig:opt}. In \Cref{fig:opt}A it can be seen that the slips that
dominate the probability vary smoothly. To explain this, note that the
map from slip patches to seafloor deformation, given by the Okada
model, is smoothing. That is, smooth slip patch patterns can lead to
similar seafloor deformations as less smooth patterns, which are
less likely in our slip patch model.
\Cref{fig:opt}C shows the tsunami waves at shore induced by
the seafloor elevation changes shown in \Cref{fig:opt}B. The tsunami
waves are shown at times $t_{\max}$ when they have maximal
average height in $[c,d]$, i.e., they maximize the right hand side in
\eqref{eq:G}. In particular the largest waves have a
steep gradient or a shock at their leading edge.

\begin{figure}
	\footnotesize
	\centering
	\begin{tikzpicture}[]
	\begin{axis}[scale only axis,
	xlabel={Observation location [km]},
	ylabel={Wave [m]}, ylabel near ticks, yticklabel pos=left,
	ymax = 7, ymin = -4,
	 ylabel shift=-2ex, 
	 xmin=30, xmax=38,
        width=.36\textwidth, height=.36\textwidth,
        axis on top,
	]
   \draw[very thick, fill=red, fill opacity=0.12, draw=none] (axis cs:34,-4) rectangle (axis cs:35,7);
   \addplot[color=black!35, line width=1, mark=none, mark size=1.5pt,]
table[x expr=\thisrowno{0}*0.001,y=wave] {\data/opt_wv7.5_z1_gamma0.03.txt};
\addplot[color=black!35, line width=1, mark=none, mark size=1.5pt,]
table[x expr=\thisrowno{0}*0.001,y=wave] {\data/opt_wv8.5_z2_gamma0.03.txt};
\addplot[color=black!35, line width=1, mark=none, mark size=1.5pt,]
table[x expr=\thisrowno{0}*0.001,y=wave] {\data/opt_wv9_z3_gamma0.03.txt};
\addplot[densely dashed, color=black!35, line width=1, mark=none, mark size=1.5pt, ]
table[x expr=\thisrowno{0}*0.001,y=wave] {\data/opt_wv9_z4_gamma0.03.txt};
\addplot[densely dotted, color=black!35, line width=1, mark=none, mark size=1.5pt,]
table[x expr=\thisrowno{0}*0.001,y=wave]
{\data/opt_wv9_z5_gamma0.03.txt};
   \addplot[color=\cOne, line width=1, mark=none, mark size=1.5pt,]
table[x expr=\thisrowno{0}*0.001,y=wave] {\data/opt_linear_wv7.5_z1_gamma0.03.txt};
\addplot[color=\cThree, line width=1, mark=none, mark size=1.5pt,]
table[x expr=\thisrowno{0}*0.001,y=wave] {\data/opt_linear_wv8.5_z2_gamma0.03.txt};
\addplot[color=\cFour, line width=1, mark=none, mark size=1.5pt,]
table[x expr=\thisrowno{0}*0.001,y=wave] {\data/opt_linear_wv9_z3_gamma0.03.txt};
\addplot[densely dashed, color=\cFour, line width=1, mark=none, mark size=1.5pt, ]
table[x expr=\thisrowno{0}*0.001,y=wave] {\data/opt_linear_wv9_z4_gamma0.03.txt};
\addplot[densely dotted, color=\cFour, line width=1, mark=none, mark
  size=1.5pt,
  name path = shock]
table[x expr=\thisrowno{0}*0.001,y=wave] {\data/opt_linear_wv9_z5_gamma0.03.txt};

\path[name path=bottom] (axis cs: 38,-4) -- (axis cs: 30,-4);     %
\addplot[cyan!50!blue!8] fill between[of=shock and bottom,soft clip={domain=30:38}];
\draw[->, very thick, -stealth] (axis cs: 37.6, -3) -- (axis cs: 35.6, -3) node[above, midway] {wave dir.};
	\end{axis}
        \end{tikzpicture}
        \caption{Comparison of waves on shore modeled by the shallow
          water equations and their linearization about the rest
          state. The waves are initialized by the bathymetry change
          corresponding to the optimizers from \Cref{fig:opt}.
          Snapshots show the waves when their average is largest in
          the observation interval. The gray waves, shown for reference, are the
          same as in
          \Cref{fig:opt}C). The waves shown in color are obtained
          by solving the linearized equations using the same
          initializations. } \label{fig:linear}
\end{figure}
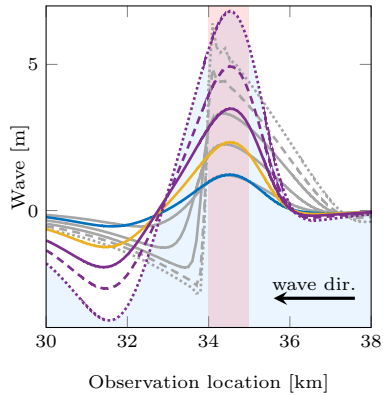

To further study this behavior, in \Cref{fig:snapshots} we show
snapshots of the wave corresponding to the $z=4$ QoI as it travels
towards shore. The wave trough and crest initially stretches over
about $80$km. As approaching shore, the wave is compressed to a few
kilometers. The wave's crest leading edge
steepens towards a shock as it approaches the observation region close
to shore. We postulate that reaching a shocked state further away from
shore would be a less efficient mechanism and, hence, is not found by the LDT-optimization.

To study this further,
in \Cref{fig:linear}, we compare snapshots of waves computed by the
shallow water equations and their linearization, both driven by the
same bathymetry change profiles shown in \Cref{fig:opt}B.  First, it
can be seen that the differences between the nonlinear and linearized
equations are more substantial for larger initial conditions, i.e.,
larger $z$. That is, the nonlinearity is more important for larger
bathymetry changes, i.e., when the linearization around the rest state
is a poorer approximation to the nonlinear problem. This observation also coincides with  results from other studies on earthquake-induced tsunamis \citep{an2014tsunami}.
 It can also  be seen
that waves from the linearized equations are generally higher and less
steep, while the waves from the shallow water equations have steep
leading edge crests, which are caused by the nonlinearity in
the equations.
Since shocks (or, correspondingly, the steep regions in our numerical
simulations in combination with the artificial viscosity) introduce
dissipation at the wave crests, in \Cref{fig:linear} the snapshots for
the linearized system are slightly higher. While the linearized
problem uses the same amount of artificial viscosity, it plays
little role as the wave's slopes are significantly smaller. This again
illustrates the importance of the nonlinearity in the equations for
the dynamics and thus the tsunami probabilities on shore.

\section{Discussion}
Using a tsunami model and a realistic distribution of random
earthquake slips, we study the probability of large tsunamis occurring
close to shore. We use an earthquake slip distribution that reflects a
realistic distribution of seismic moment magnitude from 7 to 9 on the
Gutenberg-Richter scale. We show that tsunami height probabilities on
shore can be estimated accurately by solving a series of optimization
problems and verify these probabilities by comparison with direct but
more costly Monte Carlo sampling. These optimization problems identify
the most likely and thus most effective mechanism that result in
tsunamis of a certain size. We find that the most effective earthquake
mechanisms have smoothly varying slip patches. These lead to waves
with a steepening gradient as the crest travels towards shore,
typically resulting in an approximate shock right around the region
close to shore. This indicates that the nonlinearity in the shallow
water equations, which are used to model tsunami waves, plays an
important role for the generation of tsunami waves.
We find that tsunamis with average wave height of $z\ge
3$ correspond to slips from the log-normal distribution with mean
corresponding to the $M_w=9$ earthquake.

Note that in \Cref{fig:prob}, the probability curve bends upwards.
That is, tsunami size probabilities do not decay exponentially with earthquake moment
magnitude, i.e., the Gutenberg-Richter law (see also
\Cref{fig:Mw}). Thus, model predicts a
higher rate of large tsunamis than of large earthquakes. The onset of
a similar behavior can also be seen in \citep{williamson2020source}.

A limitation of the presented results is that they are based on a
one-dimensional tsunami model and use one-dimensional slip patches to
model random bathymetry changes. From the methodology perspective, an identical approach applies to
two-dimensional shallow water models and fault slips, but would
require access to efficiently computed gradients of the
parameter-to-QoI map \eqref{eq:Ftsunami}.  Further, while the proposed
optimization-based technique provides an approximation to the true
probability, it can be combined with importance sampling to find the exact
probabilities \citep{TongVandeneijndenStadler21}. Finally, note that the
optimization-based method employed here
looses its computational advantage over Monte Carlo sampling if one
aims at computing hazard probabilities at many different points and
for many different probability thresholds $z$.

\bibliography{physic-mix}


\begin{thebibliography}{28}
\ifx \bisbn   \undefined \def \bisbn  #1{ISBN #1}\fi
\ifx \binits  \undefined \def \binits#1{#1}\fi
\ifx \bauthor  \undefined \def \bauthor#1{#1}\fi
\ifx \batitle  \undefined \def \batitle#1{#1}\fi
\ifx \bjtitle  \undefined \def \bjtitle#1{#1}\fi
\ifx \bvolume  \undefined \def \bvolume#1{\textbf{#1}}\fi
\ifx \byear  \undefined \def \byear#1{#1}\fi
\ifx \bissue  \undefined \def \bissue#1{#1}\fi
\ifx \bfpage  \undefined \def \bfpage#1{#1}\fi
\ifx \blpage  \undefined \def \blpage #1{#1}\fi
\ifx \burl  \undefined \def \burl#1{\textsf{#1}}\fi
\ifx \doiurl  \undefined \def \doiurl#1{\url{https://doi.org/#1}}\fi
\ifx \betal  \undefined \def \betal{\textit{et al.}}\fi
\ifx \binstitute  \undefined \def \binstitute#1{#1}\fi
\ifx \binstitutionaled  \undefined \def \binstitutionaled#1{#1}\fi
\ifx \bctitle  \undefined \def \bctitle#1{#1}\fi
\ifx \beditor  \undefined \def \beditor#1{#1}\fi
\ifx \bpublisher  \undefined \def \bpublisher#1{#1}\fi
\ifx \bbtitle  \undefined \def \bbtitle#1{#1}\fi
\ifx \bedition  \undefined \def \bedition#1{#1}\fi
\ifx \bseriesno  \undefined \def \bseriesno#1{#1}\fi
\ifx \blocation  \undefined \def \blocation#1{#1}\fi
\ifx \bsertitle  \undefined \def \bsertitle#1{#1}\fi
\ifx \bsnm \undefined \def \bsnm#1{#1}\fi
\ifx \bsuffix \undefined \def \bsuffix#1{#1}\fi
\ifx \bparticle \undefined \def \bparticle#1{#1}\fi
\ifx \barticle \undefined \def \barticle#1{#1}\fi
\bibcommenthead
\ifx \bconfdate \undefined \def \bconfdate #1{#1}\fi
\ifx \botherref \undefined \def \botherref #1{#1}\fi
\ifx \url \undefined \def \url#1{\textsf{#1}}\fi
\ifx \bchapter \undefined \def \bchapter#1{#1}\fi
\ifx \bbook \undefined \def \bbook#1{#1}\fi
\ifx \bcomment \undefined \def \bcomment#1{#1}\fi
\ifx \oauthor \undefined \def \oauthor#1{#1}\fi
\ifx \citeauthoryear \undefined \def \citeauthoryear#1{#1}\fi
\ifx \endbibitem  \undefined \def \endbibitem {}\fi
\ifx \bconflocation  \undefined \def \bconflocation#1{#1}\fi
\ifx \arxivurl  \undefined \def \arxivurl#1{\textsf{#1}}\fi
\csname PreBibitemsHook\endcsname

\bibitem{grezio2017probabilistic}
\begin{barticle}
\bauthor{\bsnm{Grezio}, \binits{A.}},
\bauthor{\bsnm{Babeyko}, \binits{A.}},
\bauthor{\bsnm{Baptista}, \binits{M.A.}},
\bauthor{\bsnm{Behrens}, \binits{J.}},
\bauthor{\bsnm{Costa}, \binits{A.}},
\bauthor{\bsnm{Davies}, \binits{G.}},
\bauthor{\bsnm{Geist}, \binits{E.L.}},
\bauthor{\bsnm{Glimsdal}, \binits{S.}},
\bauthor{\bsnm{Gonz{\'a}lez}, \binits{F.I.}},
\bauthor{\bsnm{Griffin}, \binits{J.}}, \betal:
\batitle{Probabilistic tsunami hazard analysis: Multiple sources and global
  applications}.
\bjtitle{Reviews of Geophysics}
\bvolume{55}(\bissue{4}),
\bfpage{1158}--\blpage{1198}
(\byear{2017})
\end{barticle}
\endbibitem

\bibitem{behrens2021probabilistic}
\begin{barticle}
\bauthor{\bsnm{Behrens}, \binits{J.}},
\bauthor{\bsnm{L{\o}vholt}, \binits{F.}},
\bauthor{\bsnm{Jalayer}, \binits{F.}},
\bauthor{\bsnm{Lorito}, \binits{S.}},
\bauthor{\bsnm{Salgado-G{\'a}lvez}, \binits{M.A.}},
\bauthor{\bsnm{S{\o}rensen}, \binits{M.}},
\bauthor{\bsnm{Abadie}, \binits{S.}},
\bauthor{\bsnm{Aguirre-Ayerbe}, \binits{I.}},
\bauthor{\bsnm{Aniel-Quiroga}, \binits{I.}},
\bauthor{\bsnm{Babeyko}, \binits{A.}}, \betal:
\batitle{Probabilistic tsunami hazard and risk analysis: a review of research
  gaps}.
\bjtitle{Frontiers in Earth Science}
\bvolume{9},
\bfpage{628772}
(\byear{2021})
\end{barticle}
\endbibitem

\bibitem{gao2018defining}
\begin{barticle}
\bauthor{\bsnm{Gao}, \binits{D.}},
\bauthor{\bsnm{Wang}, \binits{K.}},
\bauthor{\bsnm{Insua}, \binits{T.L.}},
\bauthor{\bsnm{Sypus}, \binits{M.}},
\bauthor{\bsnm{Riedel}, \binits{M.}},
\bauthor{\bsnm{Sun}, \binits{T.}}:
\batitle{Defining megathrust tsunami source scenarios for northernmost
  {C}ascadia}.
\bjtitle{Natural Hazards}
\bvolume{94}(\bissue{1}),
\bfpage{445}--\blpage{469}
(\byear{2018})
\end{barticle}
\endbibitem

\bibitem{leveque2016generating}
\begin{bchapter}
\bauthor{\bsnm{LeVeque}, \binits{R.J.}},
\bauthor{\bsnm{Waagan}, \binits{K.}},
\bauthor{\bsnm{Gonz{\'a}lez}, \binits{F.I.}},
\bauthor{\bsnm{Rim}, \binits{D.}},
\bauthor{\bsnm{Lin}, \binits{G.}}:
\bctitle{Generating random earthquake events for probabilistic tsunami hazard
  assessment}.
In: \bbtitle{Global Tsunami Science: Past and Future}
vol. \bseriesno{I},
pp. \bfpage{3671}--\blpage{3692}.
\bpublisher{Birkh\"auser},
\blocation{Cham}
(\byear{2016})
\end{bchapter}
\endbibitem

\bibitem{williamson2020source}
\begin{botherref}
\oauthor{\bsnm{Williamson}, \binits{A.L.}},
\oauthor{\bsnm{Rim}, \binits{D.}},
\oauthor{\bsnm{Adams}, \binits{L.M.}},
\oauthor{\bsnm{LeVeque}, \binits{R.J.}},
\oauthor{\bsnm{Melgar}, \binits{D.}},
\oauthor{\bsnm{Gonz{\'a}lez}, \binits{F.I.}}:
A source clustering approach for efficient inundation modeling and regional
  scale probabilistic tsunami hazard assessment.
Frontiers in Earth Science,
442
(2020)
\end{botherref}
\endbibitem

\bibitem{liu2001monte}
\begin{bbook}
\bauthor{\bsnm{Liu}, \binits{J.S.}}:
\bbtitle{{M}onte {C}arlo Strategies in Scientific Computing}
vol. \bseriesno{10}.
\bpublisher{Springer},
\blocation{New York}
(\byear{2001})
\end{bbook}
\endbibitem

\bibitem{DematteisGrafkeVandeneijnden19}
\begin{barticle}
\bauthor{\bsnm{Dematteis}, \binits{G.}},
\bauthor{\bsnm{Grafke}, \binits{T.}},
\bauthor{\bsnm{Vanden-Eijnden}, \binits{E.}}:
\batitle{Extreme event quantification in dynamical systems with random
  components}.
\bjtitle{SIAM/ASA Journal on Uncertainty Quantification}
\bvolume{7}(\bissue{3}),
\bfpage{1029}--\blpage{1059}
(\byear{2019})
\end{barticle}
\endbibitem

\bibitem{TongVandeneijndenStadler21}
\begin{barticle}
\bauthor{\bsnm{Tong}, \binits{S.}},
\bauthor{\bsnm{Vanden-Eijnden}, \binits{E.}},
\bauthor{\bsnm{Stadler}, \binits{G.}}:
\batitle{Extreme event probability estimation using {PDE}-constrained
  optimization and large deviation theory, with application to tsunamis}.
\bjtitle{Communications in Applied Mathematics and Computational Science}
\bvolume{16}(\bissue{2}),
\bfpage{181}--\blpage{225}
(\byear{2021})
\end{barticle}
\endbibitem

\bibitem{dematteis2018rogue}
\begin{barticle}
\bauthor{\bsnm{Dematteis}, \binits{G.}},
\bauthor{\bsnm{Grafke}, \binits{T.}},
\bauthor{\bsnm{Vanden-Eijnden}, \binits{E.}}:
\batitle{Rogue waves and large deviations in deep sea}.
\bjtitle{Proceedings of the National Academy of Sciences}
\bvolume{115}(\bissue{5}),
\bfpage{855}--\blpage{860}
(\byear{2018})
\end{barticle}
\endbibitem

\bibitem{rackwitz2001reliability}
\begin{barticle}
\bauthor{\bsnm{Rackwitz}, \binits{R.}}:
\batitle{Reliability analysis -- a review and some perspectives}.
\bjtitle{Structural Safety}
\bvolume{23}(\bissue{4}),
\bfpage{365}--\blpage{395}
(\byear{2001})
\end{barticle}
\endbibitem

\bibitem{HalkoMartinssonTropp11}
\begin{barticle}
\bauthor{\bsnm{Halko}, \binits{N.}},
\bauthor{\bsnm{Martinsson}, \binits{P.G.}},
\bauthor{\bsnm{Tropp}, \binits{J.A.}}:
\batitle{Finding structure with randomness: {P}robabilistic algorithms for
  constructing approximate matrix decompositions}.
\bjtitle{SIAM Review}
\bvolume{53}(\bissue{2}),
\bfpage{217}--\blpage{288}
(\byear{2011})
\end{barticle}
\endbibitem

\bibitem{FujiwaraKodairaNoEtAl11}
\begin{barticle}
\bauthor{\bsnm{Fujiwara}, \binits{T.}},
\bauthor{\bsnm{Kodaira}, \binits{S.}},
\bauthor{\bsnm{No}, \binits{T.}},
\bauthor{\bsnm{Kaiho}, \binits{Y.}},
\bauthor{\bsnm{Takahashi}, \binits{N.}},
\bauthor{\bsnm{Kaneda}, \binits{Y.}}:
\batitle{The 2011 {T}ohoku-{O}ki earthquake: Displacement reaching the trench
  axis}.
\bjtitle{Science}
\bvolume{334}(\bissue{6060}),
\bfpage{1240}--\blpage{1240}
(\byear{2011}).
\doiurl{10.1126/science.1211554}
\end{barticle}
\endbibitem

\bibitem{dao2007tsunami}
\begin{barticle}
\bauthor{\bsnm{Dao}, \binits{M.H.}},
\bauthor{\bsnm{Tkalich}, \binits{P.}}:
\batitle{Tsunami propagation -- a sensitivity study}.
\bjtitle{Natural Hazards and Earth System Sciences}
\bvolume{7}(\bissue{6}),
\bfpage{741}--\blpage{754}
(\byear{2007})
\end{barticle}
\endbibitem

\bibitem{zhan2012anomalously}
\begin{barticle}
\bauthor{\bsnm{Zhan}, \binits{Z.}},
\bauthor{\bsnm{Helmberger}, \binits{D.}},
\bauthor{\bsnm{Simons}, \binits{M.}},
\bauthor{\bsnm{Kanamori}, \binits{H.}},
\bauthor{\bsnm{Wu}, \binits{W.}},
\bauthor{\bsnm{Cubas}, \binits{N.}},
\bauthor{\bsnm{Duputel}, \binits{Z.}},
\bauthor{\bsnm{Chu}, \binits{R.}},
\bauthor{\bsnm{Tsai}, \binits{V.C.}},
\bauthor{\bsnm{Avouac}, \binits{J.-P.}}, \betal:
\batitle{Anomalously steep dips of earthquakes in the 2011 {T}ohoku-{O}ki
  source region and possible explanations}.
\bjtitle{Earth and Planetary Science Letters}
\bvolume{353},
\bfpage{121}--\blpage{133}
(\byear{2012})
\end{barticle}
\endbibitem

\bibitem{Okada85}
\begin{barticle}
\bauthor{\bsnm{Okada}, \binits{Y.}}:
\batitle{Surface deformation due to shear and tensile faults in a half-space}.
\bjtitle{Bulletin of the seismological society of America}
\bvolume{75}(\bissue{4}),
\bfpage{1135}--\blpage{1154}
(\byear{1985})
\end{barticle}
\endbibitem

\bibitem{murotani2008scaling}
\begin{barticle}
\bauthor{\bsnm{Murotani}, \binits{S.}},
\bauthor{\bsnm{Miyake}, \binits{H.}},
\bauthor{\bsnm{Koketsu}, \binits{K.}}:
\batitle{Scaling of characterized slip models for plate-boundary earthquakes}.
\bjtitle{Earth, planets and space}
\bvolume{60}(\bissue{9}),
\bfpage{987}--\blpage{991}
(\byear{2008})
\end{barticle}
\endbibitem

\bibitem{hashima2016coseismic}
\begin{barticle}
\bauthor{\bsnm{Hashima}, \binits{A.}},
\bauthor{\bsnm{Becker}, \binits{T.W.}},
\bauthor{\bsnm{Freed}, \binits{A.M.}},
\bauthor{\bsnm{Sato}, \binits{H.}},
\bauthor{\bsnm{Okaya}, \binits{D.A.}}:
\batitle{Coseismic deformation due to the 2011 {T}ohoku-oki earthquake:
  influence of 3-{D} elastic structure around japan}.
\bjtitle{Earth, Planets and Space}
\bvolume{68}(\bissue{1}),
\bfpage{1}--\blpage{15}
(\byear{2016})
\end{barticle}
\endbibitem

\bibitem{mai2002spatial}
\begin{barticle}
\bauthor{\bsnm{Mai}, \binits{P.M.}},
\bauthor{\bsnm{Beroza}, \binits{G.C.}}:
\batitle{A spatial random field model to characterize complexity in earthquake
  slip}.
\bjtitle{Journal of Geophysical Research: Solid Earth}
\bvolume{107}(\bissue{B11}),
\bfpage{10}
(\byear{2002})
\end{barticle}
\endbibitem

\bibitem{crempien2020effects}
\begin{barticle}
\bauthor{\bsnm{Crempien}, \binits{J.G.F.}},
\bauthor{\bsnm{Urrutia}, \binits{A.}},
\bauthor{\bsnm{Benavente}, \binits{R.}},
\bauthor{\bsnm{Cienfuegos}, \binits{R.}}:
\batitle{Effects of earthquake spatial slip correlation on variability of
  tsunami potential energy and intensities}.
\bjtitle{Scientific reports}
\bvolume{10}(\bissue{1}),
\bfpage{1}--\blpage{10}
(\byear{2020})
\end{barticle}
\endbibitem

\bibitem{lavallee2003stochastic}
\begin{botherref}
\oauthor{\bsnm{Lavall{\'e}e}, \binits{D.}},
\oauthor{\bsnm{Archuleta}, \binits{R.J.}}:
Stochastic modeling of slip spatial complexities for the 1979 {I}mperial
  {V}alley, {C}alifornia, earthquake.
Geophysical Research Letters
\textbf{30}(5)
(2003)
\end{botherref}
\endbibitem

\bibitem{lavallee2006stochastic}
\begin{barticle}
\bauthor{\bsnm{Lavall{\'e}e}, \binits{D.}},
\bauthor{\bsnm{Liu}, \binits{P.}},
\bauthor{\bsnm{Archuleta}, \binits{R.J.}}:
\batitle{Stochastic model of heterogeneity in earthquake slip spatial
  distributions}.
\bjtitle{Geophysical Journal International}
\bvolume{165}(\bissue{2}),
\bfpage{622}--\blpage{640}
(\byear{2006})
\end{barticle}
\endbibitem

\bibitem{kagan2013tohoku}
\begin{barticle}
\bauthor{\bsnm{Kagan}, \binits{Y.Y.}},
\bauthor{\bsnm{Jackson}, \binits{D.D.}}:
\batitle{Tohoku earthquake: A surprise?}
\bjtitle{Bulletin of the Seismological Society of America}
\bvolume{103}(\bissue{2B}),
\bfpage{1181}--\blpage{1194}
(\byear{2013})
\end{barticle}
\endbibitem

\bibitem{borzi2011computational}
\begin{bbook}
\bauthor{\bsnm{Borzi}, \binits{A.}},
\bauthor{\bsnm{Schulz}, \binits{V.}}:
\bbtitle{Computational Optimization of Systems Governed by Partial Differential
  Equations}
vol. \bseriesno{8}.
\bpublisher{SIAM},
\blocation{Philadelphia}
(\byear{2011})
\end{bbook}
\endbibitem

\bibitem{HinzePinnauUlbrichEtAl09}
\begin{bbook}
\bauthor{\bsnm{Hinze}, \binits{M.}},
\bauthor{\bsnm{Pinnau}, \binits{R.}},
\bauthor{\bsnm{Ulbrich}, \binits{M.}},
\bauthor{\bsnm{Ulbrich}, \binits{S.}}:
\bbtitle{Optimization with {PDE} Constraints}.
\bpublisher{Springer},
\blocation{Dordrecht}
(\byear{2009})
\end{bbook}
\endbibitem

\bibitem{Reyes15}
\begin{bbook}
\bauthor{\bsnm{De~Los~Reyes}, \binits{J.C.}}:
\bbtitle{Numerical {PDE}-constrained Optimization}.
\bpublisher{Springer},
\blocation{Cham}
(\byear{2015})
\end{bbook}
\endbibitem

\bibitem{hesthaven2007nodal}
\begin{bbook}
\bauthor{\bsnm{Hesthaven}, \binits{J.S.}},
\bauthor{\bsnm{Warburton}, \binits{T.}}:
\bbtitle{Nodal Discontinuous {G}alerkin Methods: Algorithms, Analysis, and
  Applications}.
\bpublisher{Springer},
\blocation{New York}
(\byear{2007})
\end{bbook}
\endbibitem

\bibitem{NocedalWright06}
\begin{bbook}
\bauthor{\bsnm{Nocedal}, \binits{J.}},
\bauthor{\bsnm{Wright}, \binits{S.J.}}:
\bbtitle{Numerical Optimization},
\bedition{2}nd edn.
\bpublisher{Springer},
\blocation{Berlin, Heidelberg, New York}
(\byear{2006})
\end{bbook}
\endbibitem

\bibitem{an2014tsunami}
\begin{barticle}
\bauthor{\bsnm{An}, \binits{C.}},
\bauthor{\bsnm{Sep{\'u}lveda}, \binits{I.}},
\bauthor{\bsnm{Liu}, \binits{P.L.-F.}}:
\batitle{Tsunami source and its validation of the 2014 {I}quique, {C}hile,
  earthquake}.
\bjtitle{Geophysical Research Letters}
\bvolume{41}(\bissue{11}),
\bfpage{3988}--\blpage{3994}
(\byear{2014})
\end{barticle}
\endbibitem

\end{thebibliography}

\end{document}